\documentclass[10pt,conference]{IEEEtran}
\IEEEoverridecommandlockouts
\usepackage{cite}
\usepackage{amsmath,amssymb,amsfonts,bm}
\usepackage{algorithmic}
\usepackage{graphicx}
\usepackage{textcomp}
\usepackage{xcolor}
\usepackage{cuted}
\usepackage[font=small,tableposition=top]{caption}
\usepackage[braket]{qcircuit}
\usepackage{lscape}
\usepackage{pgfplotstable,filecontents}
\usepackage[T1]{fontenc}
\usepackage{booktabs}
\usepackage{longtable}
\usepackage{tabu}
\usepackage{url}
\usepackage{breakurl}
\usepackage[breaklinks]{hyperref}
\usepackage{midfloat}

\hypersetup{
  breaklinks=true,
  colorlinks=true,
  linkcolor=blue,
  urlcolor=blue,
  citecolor=blue
}
\pgfplotsset{compat=1.17}
\pgfplotstableset{
    begin table=\begin{longtable},
    end table=\end{longtable},
}
\def\BibTeX{{\rm B\kern-.05em{\sc i\kern-.025em b}\kern-.08em
    T\kern-.1667em\lower.7ex\hbox{E}\kern-.125emX}}

\makeatletter
\newcommand{\linebreakand}{%
  \end{@IEEEauthorhalign}
  \hfill\mbox{}\par
  \mbox{}\hfill\begin{@IEEEauthorhalign}
}
\makeatother

\newcommand{\beq}[0]{\begin{equation}}
\newcommand{\eeq}[0]{\end{equation}}
\newcommand{\inlinecode}{\texttt}

\begin{document}

\title{A QUBO model of the RNA folding problem optimized by variational hybrid quantum annealing}

\author{

\IEEEauthorblockN{Tristan Zaborniak\textsuperscript{*}}
\IEEEauthorblockA{\textit{Department of Computer Science} \\
\textit{University of Victoria}\\
 Victoria, Canada\\
tristanz@uvic.ca}

\and

\IEEEauthorblockN{Juan Giraldo\textsuperscript{*}}
\IEEEauthorblockA{\textit{Department of Computer Science} \\
\textit{University of Victoria}\\
Victoria, Canada\\
juangiraldo@uvic.ca}

\linebreakand

\IEEEauthorblockN{Hausi Müller}
\IEEEauthorblockA{\textit{Department of Computer Science} \\
\textit{University of Victoria}\\
Victoria, Canada\\
hausi@uvic.ca}

\and

\IEEEauthorblockN{Hosna Jabbari}
\IEEEauthorblockA{\textit{Department of Computer Science} \\
\textit{University of Victoria}\\
Victoria, Canada\\
jabbari@uvic.ca}

\and

\IEEEauthorblockN{Ulrike Stege}
\IEEEauthorblockA{\textit{Department of Computer Science} \\
\textit{University of Victoria}\\
Victoria, Canada\\
ustege@uvic.ca}
}

\maketitle

\begingroup\renewcommand\thefootnote{*}
\footnotetext{These authors contributed equally to this work.}
\endgroup

\begin{abstract}

RNAs self-interact through hydrogen-bond base-pairing between nucleotides and fold into specific, stable structures that substantially govern their biochemical behaviour. Experimental characterization of these structures remains difficult, hence the desire to predict them computationally from sequence information. However, correctly predicting even the base pairs involved in the folded structure of an RNA, known as secondary structure, from its sequence using minimum free energy models is understood to be NP-hard. Classical approaches rely on heuristics or avoid considering pseudoknots in order to render this problem more tractable, with the cost of inexactness or excluding an entire class of important RNA structures. Given their prospective and demonstrable advantages in certain domains, including combinatorial optimization, quantum computing approaches by contrast have the potential to compute the full RNA folding problem while remaining more feasible and exact. Herein, we present a physically-motivated QUBO model of the RNA folding problem amenable to both quantum annealers and circuit-model quantum computers and compare the performance of this formulation versus current RNA folding QUBOs after tuning the parameters of all against known RNA structures using an approach we call ``variational hybrid quantum annealing''. 


\end{abstract}

\bigskip


\section{Introduction}

RNAs are crucial to biological function, assisting in the (de)coding and regulation of genes, protein construction, and catalysis \cite{Higgs2000, Brierley2007}. Indeed, RNA forms the genetic material in some organisms (e.g., RNA viruses) and may have preceded DNA and proteins in evolutionary history \cite{Eddy2001, Joyce2018}. Integral to these roles is the ability and tendency of RNA to fold into specific structures through hydrogen-bond base-pairing with itself \cite{halder2013rna}. These structures dictate in combination with the primary sequence the behaviour of any particular RNA and its interactions with other biomolecules \cite{conn1998rna, Holbrook2005, pan2006rna}.

While recent advances in sequencing technology have made possible rapid accessioning of RNA primary sequences, experimental determination of RNA structure remains slow and expensive by comparison \cite{Wan2011, Strobel2018, Bottaro2014, Shapiro2007, Giegerich2004}. In addition to the desire to efficiently design structured RNAs, this provides incentive toward the end of predicting RNA structure computationally from primary sequence information \cite{Westhof1996, Andronescu2004, McKeague2016, Bonnet2020}.

However, the computational approach is not without its challenges. Given that the most stable structure of a molecule is with minimum free energy (MFE) at equilibrium, many RNA structure prediction algorithms attempt to determine the MFE structure of an RNA from primary sequence information \cite{Tinoco1999}. This problem is understood to be NP-hard when allowing for all possible observed intricacies of RNA secondary structure (e.g., pseudoknots) \cite{Lyngs2000}. Various classical approaches achieve speedups through either heuristic methods or introducing simplifications to the energy model for RNA folding (e.g., ignoring pseudoknots, or focusing on only certain classes thereof) \cite{zuker1981optimal, Ruan2003, REN2005, Janssen2010, Bellaousov2010, waldsich2014rna, Takitou2019, Huang2019, Sato2021}. Other studies to optimize the balance between RNA folding speed and realism are ongoing. \cite{gray2021knotali, Gruzewski2021, https://doi.org/10.48550/arxiv.2106.07527}.

Quantum computing approaches have the potential to contribute considerably to this objective, given their proven and anticipated abilities to dramatically reduce the time-complexity of certain problem cases compared to classical counterparts, including optimization \cite{Jozsa2003, Zanca2016, Gilyn2019, Barkoutsos2020, hauke2020perspectives, Gilliam2021}. This suggests that pseudoknot-inclusive RNA folding in the MFE approximation could see improvements if framed in a manner amenable to solution by quantum computational techniques.

In particular, both quantum annealing machines and circuit-model quantum computers are capable of solving quadratic unconstrained binary optimization (QUBO) problems, while offering advantage over classical approaches \cite{Golden, Zahedinejad, Johnson2011, Boixo2014, Mukherjee2015}. Such problems have the following cost function: 

\beq
    {\cal H}_{{\rm QUBO}}(\bm{q})=\sum_i h_i q_i +\sum_{i>j}J_{ij}q_i q_j
\label{eq1}
\eeq

\noindent The parameter $h_i$ is called the bias on the $i$-th binary variable, $q_i \in \{0,1\}$, and  $J_{ij}$ is called the coupling between binary variables $i$ and $j$. The $h_i$ and $J_{ij}$ therefore define the problem to be solved. In the case of quantum annealing, the adiabatic theorem is exploited to continuously evolve the known, easy-to-prepare minimum-energy solution (ground-state) of an initial Hamiltonian to the ground-state of a final, non-commuting Hamiltonian encoding the problem cost-function \cite{Albash2018}. Circuit-model quantum computers solve QUBOs through digitized versions of adiabatic quantum evolution protocols and quantum sampling methods \cite{poulin2009sampling, temme2011quantum, farhi2014quantum, amin2018quantum}.

Given the diversity of quantum approaches suitable to solve QUBOs, a sensible challenge then for RNA folding is to design a robust energy function in QUBO form. Currently, two such QUBO functions have been presented, but each remains with various shortcomings \cite{Fox2021, Lewis2021, Lewis2021a}. In this paper, we first review these existing solutions, highlighting their respective benefits and drawbacks, before advancing a novel formulation which seeks to combine their principal virtues with additional sophistications and commenting on complexity results. We then tune the parameters of each QUBO using a training set of known RNA structures using what we term ``variational hybrid quantum annealing'', and compare their performance over an additional testing set of known RNA structures. Finally, we speculate on future directions to quantum RNA folding.


\section{Methods}

Our approach and contributions are organized into four important parts: (\ref{review}) literature review, where we discuss each current RNA QUBO turn to motivate our own work, (\ref{design}) RNA QUBO design, where we outline the features of our RNA QUBO model, (\ref{training}) RNA QUBO training, where we describe our procedure for tuning the parameters of our QUBO model against real RNA structures, and (\ref{testing}), RNA QUBO testing, where we delineate our method of testing our QUBO model against real structures using the D-Wave hybrid quantum-classical annealer.

\subsection{Literature review}\label{review}

One existing RNA QUBO model (hereafter referred to as model 1) seeks to identify the set of potential stems that maximizes both the number of consecutive base pairs and the average length of stems \cite{Fox2021}. This amounts to a minimization of the following QUBO equation:

\beq
\begin{split}
  {\cal H}_{{\rm RNA_1}}(\bm{q})&=\sum_i \{c_L(k_i-\mu)^2-c_B k_i^2\}q_i \\
  &-\sum_{i>j}\{2c_B k_i k_j \delta_{ij} + \delta_{ij}^{\prime}\}q_i q_j
\end{split}
\label{eq2}
\eeq

\begin{figure*}
    \centerline{\includegraphics[width=18cm]{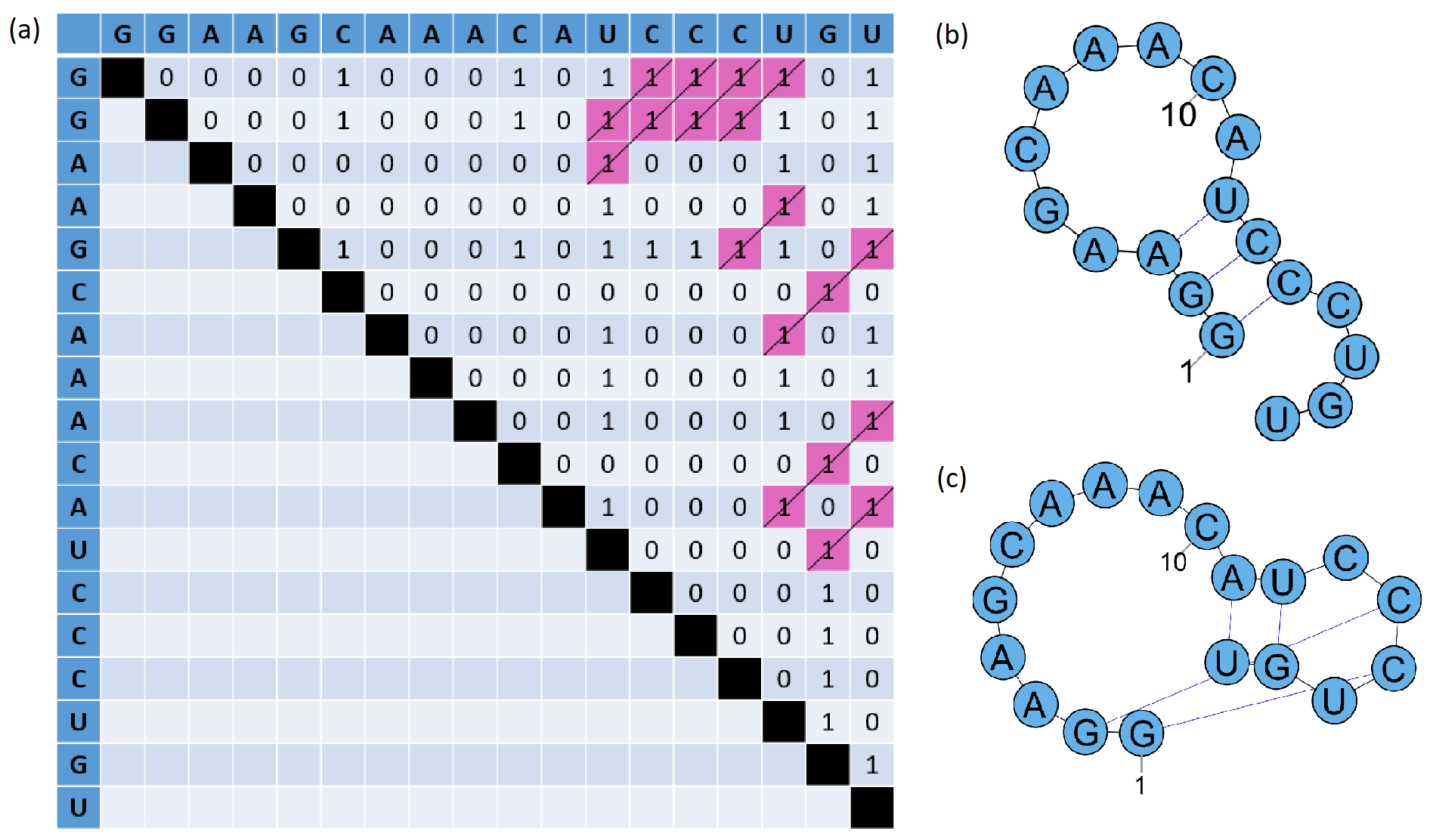}}
    \caption{(a) Matrix representation of potential base-pairs and stems for an example sequence. Potential base-pairs are marked with a 1, and stems (of at least two successive base-pairs) are marked in pink with diagonal lines to show the base-pairs involved in the stem. Note that any stem of length 3 or greater consists of overlapping stacked quartets (stems of length 2). For model 2, stacked quartets are mapped to individual qubits; for models 1 and 3, stems are mapped to individual qubits. (b) A possible structure for the example sequence comprised of one stem of length 3. In this case, two stacked quartets combine to form this stem. (c) A possible structure for the example sequence comprised of two stems, each of length 2. Notice that these stems are pseudoknotted. This figure is adapted from ref. \cite{Fox2021}.}
    \label{fig1}
\end{figure*}

\noindent Here, $q_i$ corresponds to the binary variable for the $i$-th potential stem, the parameters $c_L$ and $c_B$ are tunable constants, $k_i$ is the length of potential stem $i$ measured in base-pairs, and $\mu$ is the longest stem length in the potential set, all determined by classical pre-processing. A stem is a set of consecutive base-pairs, as shown in Figure \ref{fig1}. $\delta_{ij}$ and $\delta_{ij}^{\prime}$ are delta-function penalties applied when $i$ and $j$ are pseudoknots and overlapping, respectively. They are defined as:

\begin{subequations}
\begin{align}
\delta_{ij}&=
\begin{cases}
  1 & \text{if $i$, $j$ are not pseudoknotted} \\
  X & \text{if $i$, $j$ are pseudoknotted}
\end{cases}
\label{eq3a}
\\
\delta_{ij}^{\prime}&=
\begin{cases}
  1 & \text{if $i$, $j$ are not overlapping} \\
  -\infty & \text{if $i$, $j$ are overlapping}
\end{cases}
\label{eq3b}
\end{align}
\end{subequations}

\noindent Note that $\delta_{ij}$ increasingly penalizes pseudoknots with decreasing $X$. The authors of \cite{Fox2021} chose $X = 0.5$ in equation (\ref{eq3a}).

The linear terms of the QUBO model consist of two parts, one (associated with $c_L$) comparing the length of the $i$-th stem to the longest stem of the potential set, thereby enforcing the notion that average stem length should be maximized, and the second (associated with $c_B$) offering quadratic reward to longer stems. The quadratic terms penalize pseudoknotted stems in proportion to the product of their lengths ($\delta_{ij}$), and prevent overlapping stems from appearing in the solution based on an overwhelming constant addition to the energy function if this is the case ($\delta^{\prime}_{ij}$).

Though the essence of this QUBO model reflects certain known features of RNA folding (that longer stems are more energetically favorable, and that pseudoknotted stems are less stable than the same stems not in pseudoknot), these reflections are fairly coarse-grained. For one, the length of a stem is not directly proportional to its stability. The various typical base pairs $(G, C)$, $(A, U)$, and $(G, U)$ have differential stabilities when considered independently \cite{Varani2000} that combine non-trivially in nearest-neighbor stacking interactions \cite{xia1998thermodynamic, andronescu2014determination}. Furthermore, the approximation that pseudoknots should be penalized in proportion to the product of their lengths does not match previous, experimentally-informed attempts at representing these structures and remains without strong physical reasoning \cite{dirks2004algorithm, andronescu2010improved, Hajdin2013}. Finally, given that this QUBO is heuristic in its construction, optimal performance must require parameter tuning and testing against known RNA structures; to the best of our knowledge, this has not been carried out so far.

It should be noted that each term of equation \ref{eq2} must be pre-computed classically before the Hamiltonian can be embedded within quantum hardware for solving. It is therefore important to understand the computational complexity involved for reference when comparing to existing, classical approaches. The appendix offers proof that the time-complexity of the pre-computation step for this Hamiltonian is $\mathcal{O}(N^3\log N)$, where $N$ is RNA length, and the length of the binary vector $\bm{q} = S(N, m)$, where $m$ is the minimum stem length, is $\mathcal{O}(N^3)$ in the worst case. Native RNAs, however, typically demonstrate a $\bm{q}$-length on the order of $\mathcal{O}(N^2)$ (computed based on a fit of $S(N, m) = aN^b$ to all structures of the Archive II RNA benchmarking suite \cite{Sloma2016}).

\smallskip
\smallskip

The other existing RNA QUBO model (hereafter referred to as model 2) seeks to identify the set of potential stems of length two (stacked quartets) that maximizes successive stacked quartets while guarding against overlaps and penalizing pseudoknots \cite{Lewis2021, Lewis2021a}. This amounts to a minimization of the following QUBO equation:

\beq
  {\cal H}_{{\rm RNA_2}}(\bm{q})=-\sum_i N_iq_i -\sum_{i>j}M_{ij}q_i q_j
\label{eq4}
\eeq

\noindent Here, $q_i$ is the binary variable corresponding to the $i$-th potential stacked quartet, $N_i$ is the nearest-neighbor stacking energy of the $i$-th quartet determined from optical melting experiments at 37 $^{\circ}$C \cite{Turner2009}, and $M_{ij}$ is of the form:

\beq
M_{ij}=
\begin{cases}
  M^+ & \text{if $i$, $j$ are stacked} \\
  M^- & \text{if $i$, $j$ are pseudoknotted} \\
  M^P = -\infty & \text{if $i$, $j$ are overlapping}
\end{cases}
\label{eq5}
\eeq

\noindent $M^+$ rewards consecutive stacked quartets to incentivize stem formation, $M^-$ penalizes pseudoknot formation between stacked quartets, and $M^P$ is a large penalty that prevents the same base from being assigned to a different matching pair. $M^+$ and $M^-$ are positive, tunable values, with $M^- < M^+$ representing a soft constraint, where violation of the constraint might allow for the formation of a longer stem that further minimizes the cost-function. $M^P$, by contrast, is negative.

An advantage of this approach versus the former is its direct use of experimental data to inform the model, where the real free energy contributions of stacked quartets are considered. However, due to the computational variables being defined at this level, it is not possible to reward the formation of larger stems containing some number of base pairs versus several smaller stems totalling the same number of base pairs. Further, as with the previous model, the pseudoknot penalty is again proportional to the product of the stem lengths forming the pseudoknot, which deviates significantly from previous work \cite{dirks2004algorithm, andronescu2010improved, Hajdin2013}. Finally, the $M_{ij}$ terms are not experimentally-informed, and so must be pre-selected in advance, or tuned against known RNA structures. It is our understanding that this tuning has not been carried out previously.

As with the previous model, each parameter must be pre-computed classically before quantum embedding. The computational complexity of this stage is shown in the appendix to be on the order of $\mathcal{O}(N^2\log N)$ in time, and generates a binary vector $\bm{q}$ of $\mathcal{O}(N^2)$ in length. Therefore, there is a polynomial advantage to this approach versus the former in pre-processing time and quantum space-complexity, which becomes significant at larger sequence lengths.

\subsection{RNA QUBO design}\label{design}

\begin{figure}
    \includegraphics[width=9cm]{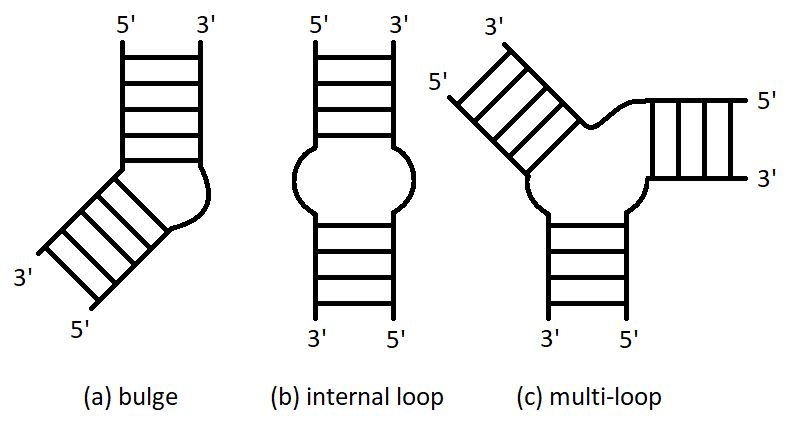}
    \caption{RNA structural motifs that are difficult to penalize directly within an RNA folding QUBO model. (a) bulges consist in unpaired stretches of base located within one strand of an otherwise continuous``stem", (b) internal loops consist in unpaired stretches of base separating two stems, and (c) multi-loops are internal loops with three or more participant stems.}
    \label{fig__motif}
\end{figure}

As outlined in section \ref{review}, we identify several features desirable at this stage in an RNA QUBO model, some of which are implemented variously by the two existing solutions covered. These features include: physically-informed representations of base-pairing energies at the stem level and physically-informed pseudoknot penalties. Further, we consider rewarding the formation of larger stems over a number of smaller stems that are the same in summed energy when considering only nearest-neighbor interactions important within a QUBO model context. Specifically, this originates from the fact that bulges, internal loops, multi-loops, and other such structural motifs are known to be destabilizing \cite{nowakowski1997rna}, features not feasibly pre-computable given that we must enumerate every possibility in advance prior to submission to a QUBO. See Figure \ref{fig__motif} for a schematic representation of these structures. This heuristic therefore favors the formation of continuous base-pairs, and may be constructed as to be tunable, as per ref. \cite{Fox2021}.

With these considerations in mind, a sketch of our proposed QUBO model (hereafter called model 3) is as follows: we first pre-compute and optimize over stems as opposed to stacked quartets, even though this comes with a slight polynomial cost. This permits us to appraise in advance the lengths of two stems that may be involved in a pseudoknot, thereby allowing for pseudoknots to be penalized in a more sophisticated manner. Moreover, computing stems allows for an understanding of possible hairpin loop lengths, and the inclusion of an associated penalty. Secondly, we opted to maintain the term of ref. \cite{Fox2021} which compares the base-pair length of the $i$-th potential stem to the maximum potential stem length in order to maximize the average stem length. Here, we replace base-pair stem length with total nearest-neighbor-based stem energy. Finally, we include a term that is linear in nearest-neighbor stem energy, and incorporate a penalty which prevents against bases from being forced to pair un-physically with more than one base.

\begin{figure}
    \includegraphics[width=9cm]{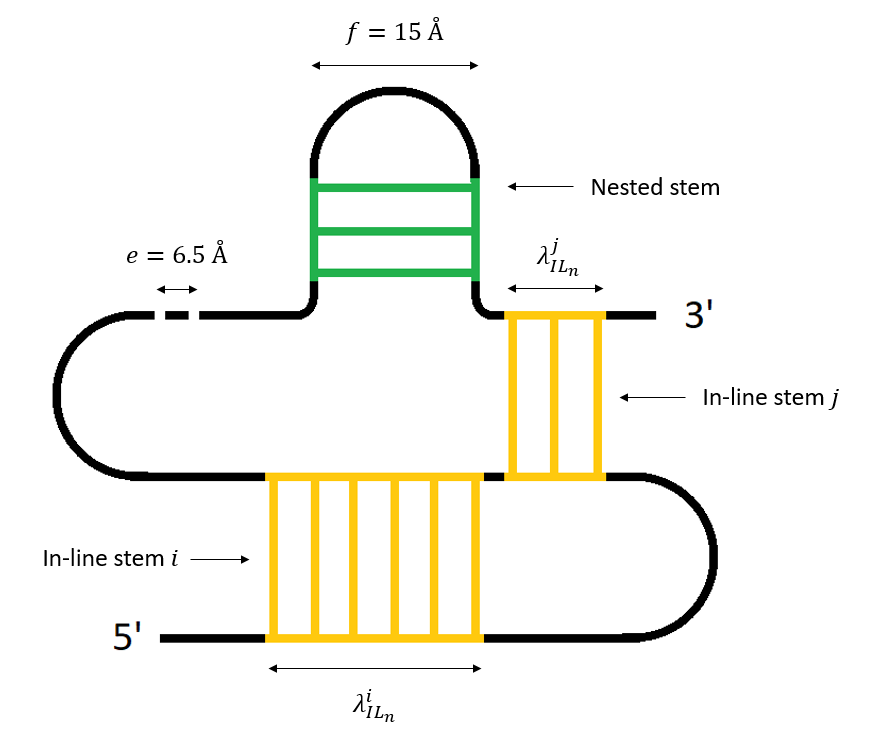}
    \caption{Overview of pseudoknot structure model and entropic penalty terms. Energy penalties for single-stranded nucleotides and nested stems are based on ref. \cite{Aalberts2010}, and the penalty for in-line stems was developed in Hajdin et al., 2013 \cite{Hajdin2013}. This figure is adapted from ref. \cite{Hajdin2013}.}
    \label{fig2}
\end{figure}

The pseudoknot penalty we abstract from ref. \cite{Hajdin2013}, which is based on a polymer-physics model \cite{Aalberts2010}. The form of this penalty is based on the insight that pseudoknots typically exhibit short single-stranded sections between their constituent in-line stems, and avoid the inclusion of nested stems. Essentially, the energetic penalties associated with single-stranded nucleotides and nested stems between in-line stems are in proportion to the linear dimension of each of these features along the strand, as shown in Figure \ref{fig2}. Furthermore, in-line stems are penalized based on their length with reference to the distribution of in-line stem lengths of known structures (see supplementary material of ref. \cite{Hajdin2013}). In full form, this penalty ($P_{PK}$) appears as:

\beq
\begin{split}
    P_{PK} &= P_1 \ln(e^2 N_{SS} + f^2 N_{NH}) \\
    & + P_2 \ln \Big(\sum_k \lambda_{IL_n}^k\Big)
\end{split}
    \label{eq6}
\eeq

\noindent Where $P_1$ and $P_2$ are adjustable parameters, $e$ and $f$ are the linear dimensions of nucleotides and nested stems in-strand, respectively, $N_{SS}$ and $N_{NH}$ are the numbers of single-stranded nucleotides and nested stems involved in the pseudoknot, respectively, and $\lambda_{IL_n}^k$ is the penalty constant for the $k$-th in-line stem of length $n$.

Now, as before with internal loops and bulges, it is difficult with the QUBO approach to feasibly pre-compute this penalty term per pair of stems participating in a pseudoknot, given that many potential nested stems may be involved in the actual pseudoknot. As a compromise, we forgo the $N_{NH}$ term and simply consider the number of nucleotides between the constituent stems, assuming them to be single-stranded, such that we operate with this modified pseudoknot penalty:

\beq
    P_{PK}^{\prime} = P_1 \ln(e^2 N_{SS}) + P_2 \ln \Big(\sum_k \lambda_{IL_n}^k\Big)
    \label{eq7}
\eeq

\noindent Assuming the previous model to be more accurate, all else equal, this compromise has us over-penalizing pseudoknots with large loops to some degree. Nevertheless, the essence of the model is captured in that larger loop structures are penalized more within the context of a pseudoknot.

We also incorporate a hairpin loop penalty ($l_i$) when considering a single stem, as per refs. \cite{Turner2009, Kelley2018-fc}. With this penalty, loops of size 1 and 2 are forbidden, while loops of size 3 or greater are with decreasing penalties until loops of size 7, beyond which all loops are assigned the same penalty. It is possible that the nucleotides of certain hairpin loops penalized in this way might participate in pseudoknots, thereby negating the need for a loop penalty (given that single-stranded sections to pseudoknots are penalized according to equation \ref{eq7}), but the probability of pseudoknots forming with short hairpin loops, where the loop penalty is highest in magnitude and rate of change, is small. Finally, we include a linear term proportional to the nearest-neighbor energy of the $i$-th stem ($k_i$) and the delta-function of equation \ref{eq3b}, yielding this QUBO equation:

\beq
\begin{split}
  H&=\sum_i \{\alpha(k_i-\mu)^2-\beta(k_i-l_i)\}q_i\\
  &+\sum_{i, j>i}(P_{PK}^{\prime}+\delta_{ij}^{\prime})q_iq_j
\end{split}
\label{eq8}
\eeq

\noindent  $\alpha$ and $\beta$ of equation \ref{eq8} are tunable parameters, as are $P_1$ and $P_2$ in equation \ref{eq7}. In section \ref{training}, we describe our tuning of these parameters, as well as those which appear in the existing solutions of section \ref{review}.

\subsection{RNA QUBO training}\label{training}

\begin{figure*}
    \centerline{\includegraphics[width=18cm]{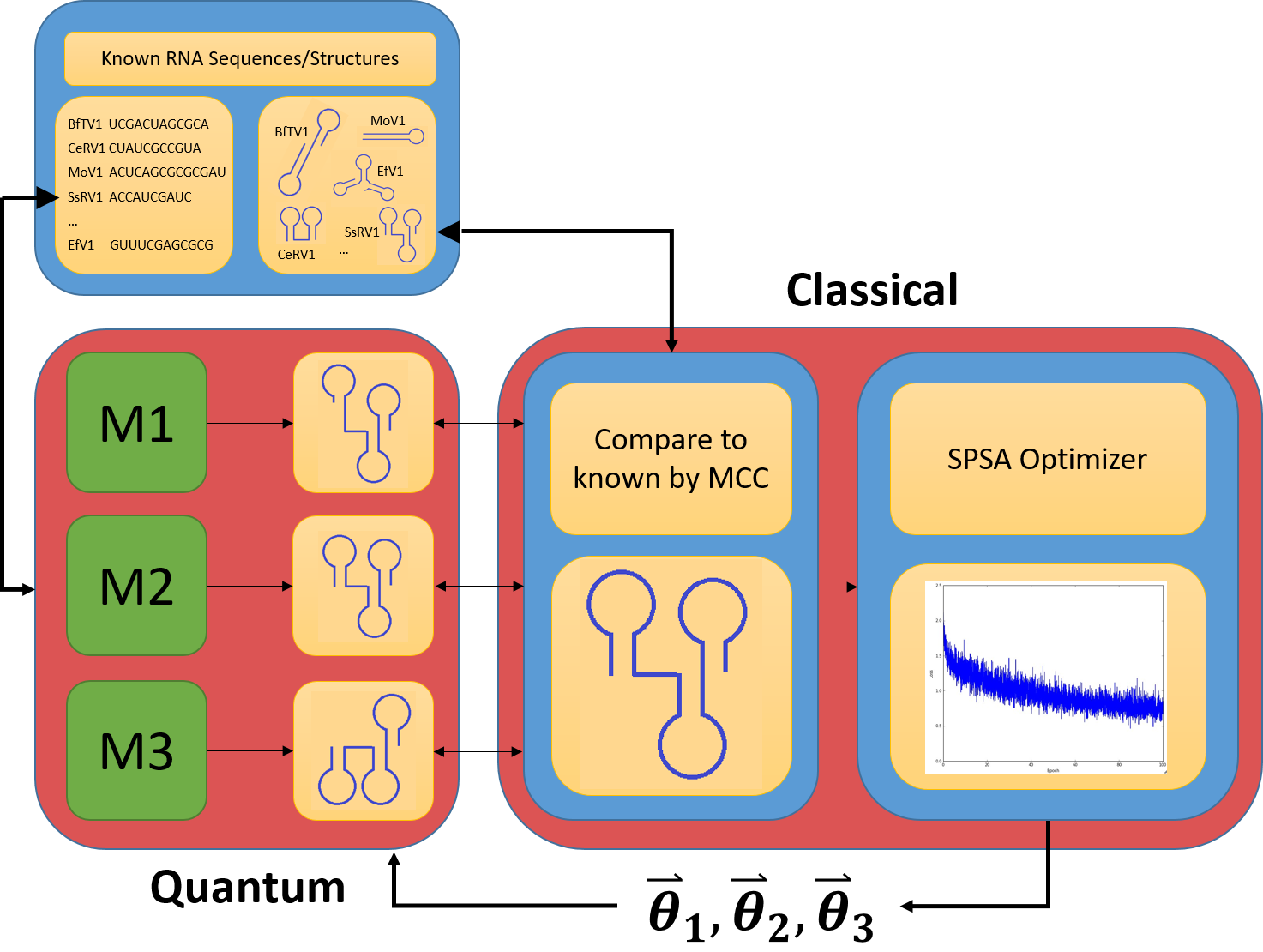}}
    \caption{Overview of RNA QUBO training protocol. RNA sequences with known structures are passed to parameterized QUBO models, which are executed by the D-Wave hybrid solver. Predicted structures of each model are compared to known structures using the MCC scoring function \cite{Matthews1975}. An SPSA optimizer iteratively updates the parameters of the QUBO models. The arithmetic mean of the MCC evaluation metric over all test structures is optimized.}
    \label{fig3}
\end{figure*}

To enable the evaluation of our parameterized RNA folding QUBO model against the models of previous work and to assess general performance, we tuned the parameters of each model by training over known RNA secondary structures with and without pseudoknots. To accomplish this, we collected 110 sequences and associated connectivity table structure files with less than 90\% sequence identity from the bpRNA-1m database \cite{Danaee2018} ranging between 20 and 50 bases in length. We divided the RNAs into training and testing groups, containing 70 and 40 sequences, respectively. Within each of these groups, half contained pseudoknots. The purpose of the with/without pseudoknots division is in following the RNA structure prediction benchmarking practices outlined in ref. \cite{Mathews2019}, where it is recommended that structures with and without pseudoknots be tested against models separately and their individual performances noted. Training was carried out using the D-Wave Advantage system together with a classical parameter optimizer as instantiated in the Amazon Braket Hybrid Jobs platform. Below we present our approach in more detail, which is summarized in Figure \ref{fig3}. We propose to call this approach ``variational hybrid quantum annealing'', in parallel to gate-based variational hybrid quantum algorithms.

\textbf{Stem and stacked-quartet identification:} We implemented two versions of an algorithm to extract lists of potential stems and stacked quartets per sequence, respectively. A potential base-pairing matrix populated only in the upper diagonal as shown in Figure \ref{fig1} was constructed per RNA sequence, where 0s indicate the impossibility of a base-pair, and 1s base-pair possibilities. We considered $(G, C)$, $(A, U)$, and $(G, U)$ as legitimate base-pairs. We proceeded to extract stems (of at least two base-pairs, given the basis of model 3 being in nearest-neighbor stacking interaction energies) and stacked-quartets by scanning matrices for consecutive possible base-pairs in upper-right diagonals in $\mathcal{O}(N^2)$ time. Following ref. \cite{Fox2021}, we represented stems and stacked quartets by listing the indices of the first and last base involved in pairing, and either the length of the stem (in the case of model 1) or the total nearest-neighbor stacking interaction energy (in the case of model 2 and model 3), keeping track of the largest potential stem, $\mu$, in the case of model 1 and model 3. The stem identification process for an example sequence \inlinecode{5'-GGAAGCAAACAUCCCUGU-3'} is depicted in Figure \ref{fig1}.

\textbf{Pseudoknot and overlap penalties:} For each model and RNA considered, each pairwise combination of stems or stacked quartets was checked for pseudoknotting and overlap using a line-sweep interval overlap algorithm. For pseudoknots, we created lists of 3-tuples that identified a combination of stems $(i,j)$ and their penalty score. If stems were found to be in pseudoknot in the case of models 1 and 2, this penalty score assumes the form of a tunable constant. With model 3, each potential pseudoknot was evaluated according to equation \ref{eq7}, requiring an evaluation of the assumed single-stranded nucleotides within the pseudoknot and a constant lookup of $\lambda_{IL_n}$ with respect to the lengths of the participant stems. In the case of overlapping stems, we used the same 3-tuple notation, but with the constant penalty $-\infty \approx -1000$.

\textbf{Parsing training structure data:} As mentioned at the beginning of this section, we retrieved the known secondary structures for each sequence as connectivity table files from bpRNA-1m. We implemented an auxiliary method to take the contents of a connectivity table file and output the stems in a format matching the potential stems on a per-model basis, as well as the length of the longest actual stem, $\mu$, in model 1 and model 3. This allowed for easy comparison of the known-structure base-pairs of a sequence to the base-pairs predicted by the models, as well as energy evaluation of known structures under the assumption of the models. 

\textbf{Model construction:} With lists of all possible stems or stacked quartets and pairwise combination penalties for a sequence, we employed equations \ref{eq2}, \ref{eq4}, and \ref{eq8} to instantiate each QUBO model programmatically in Python dictionary form, one dictionary corresponding to linear terms and the other dictionary to quadratic terms. The keys of the former we set to individual stems or stacked quartets, and the latter we set to pairs of stems or stacked quartets. To start, we initialized all tunable parameters of each model to 1.

\textbf{Hybrid quantum annealing:} Robust quantum error correcting codes are yet to be implemented on quantum annealing machines, such that purely quantum processing unit-based (QPU-based) outputs are noise-prone. For this reason, ref. \cite{Fox2021} executed their experiment on the D-Wave hybrid solver, which breaks a QUBO into smaller sub-QUBOs and iterates between running these sub-QUBOs on the D-Wave QPU and through a classical tabu search after recombination \cite{qbsolv}. We opted for this same hybrid approach using the 5000-qubit D-Wave Advantage machine with Pegasus-graph qubit connectivity, keeping default settings to the annealing schedule. As detailed above, we formulated each problem corresponding to an RNA sequence, model, and pseudoknot penalty in a dictionary form acceptable by the D-Wave QPU interface. Problems are mapped such that single stems or stacked quartets correspond to individual qubits, and the couplings between any pair $(i, j)$ map to chains of interconnected qubits between $(i, j)$. For each run, the solution was returned as a binary vector with a length corresponding to the number of potential stems or stacked quartet of the tested RNA, entries of 1 recording predicted stems or stacked quartets, and an entries of 0 recording non-predicted stems or stacked quartets. Additionally, the energy of the predicted solution and minimum estimated time-to-solution were returned per run.

\textbf{Model evaluation}: To evaluate the predicted structures versus the known structures of an RNA, the Matthews Correlation Coefficient (MCC) score over exact base-pairs was employed \cite{Matthews1975}. This metric measures the quality of binary classifications, taking into account true positive, true negative, false positive, and false negative counts such that perfect predictions score $+1$, total disagreement is indicated by $-1$, and random predictions score $0$. Versus other two-class confusion matrix scoring metrics such as $F_1$, accuracy, markedness, bookmaker informedness, and balanced accuracy, the MCC in general proves more informative and reliable \cite{Chicco2021}. Our requirement that base-pairs must be exact excludes similar solutions from scoring well \cite{Mathews2019}. For example, if we consider an RNA with the single known stem $(4, 13, 3)$, where the last tuple entry records the base-pair length of the stem, the predicted stem $(5, 14, 3)$ will score poorly, even though four of the six bases involved in pairing in the known structure are paired in the predicted. However, strict training is enforced through use of this scoring protocol.

\textbf{Model training}: We trained the parameters of each model over the 30 training structures using the Qiskit implementation of the simultaneous perturbation stochastic approximation (SPSA) optimization algorithm while carrying out QUBO function evaluations using the D-Wave hybrid solver as described \cite{Spall1992, Qiskit}. This method is based on a gradient approximation using two measurements of the loss function, where we took the arithmetic mean of the MCC score across all structures of a given training class is taken to be the loss, and perturbations to the current parameter set are stochastically generated from a zero-mean Bernoulli distribution. We observed convergence of this optimization at approximately 30-60 iterations of the optimizer in each model implementation.


\subsection{RNA QUBO testing}\label{testing}

Machine learning methods run the risk of over-fitting to training data \cite{dietterich1995overfitting, roelofs2019meta, ying2019overview}. It is therefore necessary to test machine-learning models against data other than that used in training to evaluate their realism and applicability to future instances. We test each fit model as described in (\ref{training}) over 40 known RNA structures with less than 90\% sequence identity versus those used in training, divided as before into two groups: one containing 20 structures with pseudoknots, and the other containing 20 structures without pseudoknots. We test for significant difference in performance between models over all test structures (with and without pseudoknots) using the Kolmogorov-Smirnov 2-sample test, and additionally test for intra-model difference in performance between structures with and without pseudoknots.

\section{Results}

Training each RNA folding QUBO yielded a set of parameters optimized against the training set, as well as a mean-squared loss score characterizing goodness of fit. For model 1, we found the parameters: $c_L=0.639$, $c_B=0.223$, and $\delta=0.681$, with a loss of $0.113$. For model 2, we found: $M^+ = 1.748$ and $M^- = 0.386$, with a loss of $0.263$. For model 3, we found: $\alpha = 1.604$, $\beta = 2.212$,  $P_1 = 1.495$, and $P_2 = 1.338$, with a loss of $0.116$.

\begin{figure}
    \includegraphics[width=9cm]{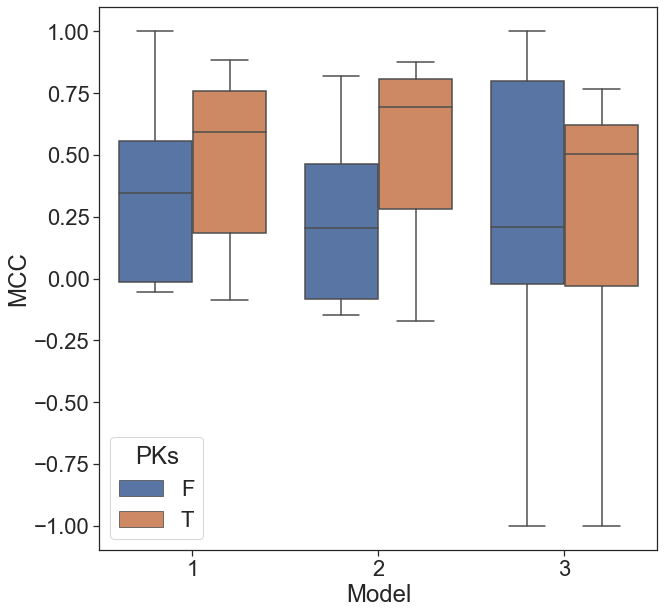}
    \caption{Performance of RNA folding QUBOs after SPSA training as measured by the Matthews Correlation Coefficient. The same 40 structures (20 structures with pseudoknots, labelled `T' and 20 structures without pseudoknots, labelled `F') were evaluated per model. Whiskers extend to the the lowest (highest) data point within 1.5 times the interquartile range as measured from the lower (upper) quartile.}
    \label{fig4}
\end{figure}

Given these optimized parameters, we tested each model against an additional 40 RNA structures. We present the results of this testing in Figure \ref{fig4} as a boxplot with reference to MCC, where we differentiate performance over structures with pseudoknots and structures without pseudoknots. Comparing model performance over all test structures using the Kolmogorov-Smirnov 2-sample test, we did not find any significant difference between them (at significance levels of $0.01$ and $0.05$). However, it should be pointed out that models 1 and 2 demonstrate a tighter variance than model 3.

Additionally, it may be seen that models 1 and 2 appear to differ appreciably in their performance over structures with pseudoknots versus structures without. A Kolmogorov-Smirnov 2-sample test confirms this difference at both $0.01$ and $0.05$ levels of significance, while reporting model 3 as consistent in its performance across structures with and without pseudoknots.

\section{Discussion}

We approach this section by first discussing our results and contributions in further detail, before proceeding to the limitations of our methodology and suggestions about how it might be improved.

\subsection{Performance of models over pseudoknotted versus unpseudoknotted structures}

We begin by considering the significant difference in performance seen in models 1 and 2 over structures with pseudoknots versus structures without pseudoknots. We suspect that this is due to our training and test structures being limited to sequences of less than 50 bases. With short sequences, the number of potential stems is quite small, and the number of potential stems in potential pseudoknots of course even smaller. Therefore, it is likely that if a potential pseudoknot (or multiple) is present for a given sequence, the actual structure will contain a pseudoknot. This means that during training, it is possible for the pseudoknot penalty to offer a relatively high reward for pseudoknot formation without compromising too much model performance over structures without pseudoknots. In model 3, despite no statistically significant difference, the median MCC score of structures with pseudoknots versus those without is greater by an appreciable degree.

These non-pseudoknotted structures are at somewhat a ``disadvantage'' in that they are with one less term in helping to determine which of a set of possible structures is correct. Consequently, it is more difficult to select the proper structure, as the energy landscape for such structures is of a higher degeneracy than structures with pseudoknots in the small regime we have tested. We expect that the difference in performance between structures with and without pseudoknots should lessen in each model if trained over larger RNAs, but this remains to be tested. We suggest this step as necessary, as it should be the case that the trained parameters reflect a training set representative of the full array of RNA structures. In this work, we limited ourselves to small structures to expedite computations.

\subsection{Comparison of overall model performance}

We now comment on the negligible difference in overall performance between the models reviewed, as determined by the Kolmogorov-Smirnov 2-sample test. This result indicates that with respect to the training and testing structures of this work, model 3 does not demonstrate any advantage versus models 1 and 2, representing existing RNA folding QUBO models. This is despite deliberate construction of model 3 to account for terms excluded from consideration by models 1 and 2 that are commonly included within the majority of other RNA folding models. Specifically, model 3 includes hairpin loop penalties and a pseudoknot penalty which takes into account intervening bases involved in a pseudoknot of two stems (either single-stranded or in nested stems). 

\subsection{Limitations of approach}

Before speculating further, we discuss limitations that constrain the presented results. One, already noted, concerns the number and length of RNA structures used in both training and testing. Due to a limited amount of computational resources and the larger amount of time it takes to compute longer RNA sequences using the D-Wave hybrid solver, we had to restrict our training data set to only 30 small structures (from 20 to 50 bases). This might affect the ability of the models to capture the effect of structure size on their performance. Furthermore, the SPSA optimizer used during the training phase is based on gradient descent, which can converge on local minima for certain starting points. We acknowledge this being a possibility within the training of our models and suggest to try different starting points for the optimizer in further iterations of this work, or a Monte Carlo optimization approach to avoid local minima trapping. We note as well that the D-Wave hybrid solver is not guaranteed to find the global optimum of a given problem instance. Typically, many executions of the same problem instance are computed to determine the optimum with greater likelihood, but due to resource limitations we opted for a greedy approach in which only one execution of each problem instance was run, both in the training and testing phases of our methods.  

Turning our attention once again to specific shortcomings of model 3, we first remark that its pseudoknot penalty modeled on that of Hajdin et al. \cite{Hajdin2013} ignores the possibility of nested stems, assuming and penalizing instead the intervening bases as if they are all single-stranded. Additionally, our hairpin loop penalty assumes the stem involved is in fact a hairpin loop, with the loop not being involved in another stem. This is clearly impossible to satisfy in every case, especially as structure size increases. It is true as well that we do not consider direct penalizations of internal loops, multi-loops, bulges, and other such structural motifs, except through the heuristic that tends to avoid them in favour of longer stems if possible. We note that some of these penalties may prove beyond the capacity of the QUBO representation: multi-loops, for instance, are formed more than by pairwise interactions between stems. 

Indeed, in general it might prove exceedingly challenging to design a sufficiently accurate RNA folding QUBO model for the reason that the method essentially demands the encoding of all possible structures at once. To achieve this while penalizing common RNA structural motifs that do not lend themselves easily to pairwise interaction terms in advance of solving the QUBO will likely remain possible only through clever heuristics. For example, consider an internal loop. If modelling at the stem level, we can determine whether two stems form an internal loop together. However, it is not possible to penalize this internal loop directly, as it might be that a portion of the internal loop will actually prove involved in another stem. However, it might be possible to penalize internal loops directly if they are smaller than a certain size, as beyond this size, it becomes impossible (or at least very improbable) for the internal loop to participate in other structures. A similar approach might be used for hairpin loops and pseudoknots, perhaps informed by empirically-derived statistical models of the probabilities of their being of a certain size, but complex motifs such as kissing hairpins and multi-loops escape simple treatment.

Despite these difficulties, we suggest the pursuit of a QUBO model of the RNA folding problem as one of salience, as this method lends itself well to the two leading quantum computing architectures (annealing and circuit-model), both of which may result in quantum advantage versus classical machines in solving QUBOs. Though in our work we execute RNA folding QUBOs by way of annealing hardware only, we emphasize that these models are easily transferable to universal quantum computers, where algorithms such as the Variational Quantum Eigensolver (VQE) and the Quantum Approximate Optimization Algorithm (QAOA) provide a feasible way of computing solutions to QUBOs.

\subsection{Variational hybrid quantum annealing}

In this section, we highlight the uniqueness of our method. We present a practical example of how quantum annealing can be used in the context of a supervised machine learning problem, where our RNA folding QUBO models are the analogs to ML models that then were trained using a hybrid quantum-classical solver in coupling with a classical gradient descent optimizer. To the best of our knowledge, this is the first experimental paper that demonstrates the implementation of variational hybrid quantum annealing. Other research presents different approaches to exploiting quantum annealing in machine learning, such as using it to optimize the training pipeline of a classical ML model for classification problems \cite{Nath2021} or to construct diabatic Ising-Born machines that act as the analog of an ansatz for parameterized quantum circuits in generative learning problems \cite{Crosson2021}. Given the similarity in structure of variational quantum algorithms on circuit-model hardware, we suggest our method as the quantum annealing extension to these methods. We also suggest that though the classical optimization of this work took place at a large physical distance from the QPU, speedup might be seen if classical resources dedicated to optimization were to be integrated in close proximity to the QPU, as with Qiskit Runtime \cite{johnson2022qiskit}. 

\subsection{Future work}

Moving forward, we expect to include larger and more complex RNA structures in the training process of the models to identify possible size dependence of the optimized model parameters. As well, we plan to increase the number of optimizer iterations and initial parameter points used to train the models to ensure that the final set of parameters represents the lowest loss possible. Another important consideration is to thoroughly review the structures which performed particularly poorly in model 3, and seek to improve the corresponding Hamiltonian based on our findings. We also plan to investigate penalizations of other structural motifs besides hairpin loops and pseudoknots as mentioned. One final point is that MCC does not provide a full measure of fitness for an RNA folding model. For instance, the relative difference in energy between the predicted and actual structures of a large suite of test sequences under a given model suggests the closeness of this model to reality if the model is reasonably complex, where this metric tends to zero as the model converges to accurate replication of reality. We aim to consider this means and others in assessing model performance.

\section{Conclusion}

We present herein a novel, physically-motivated RNA folding QUBO model amenable to quantum computers, and train the free parameters of this and other existing RNA folding QUBOs using hybrid quantum-classical annealing. We find that after training, all models perform similarly, but stress that both training and testing were limited to RNA sequences of less than 50 bases, and that working with larger sequences will be necessary before full conclusions of relative performance between models can be drawn.

We would like to highlight our approach of combining a quantum annealing solver with classical parameter optimization as an additional important contribution of this paper. While existing quantum variational algorithms employ this principle with circuit-model solvers, to date it appears that its application to quantum annealers has not been carried out. We suggest that our variational hybrid quantum annealing approach might be of especial benefit to training the free parameters of Ising problem models.

Finally, despite the limitations of RNA folding QUBO models (i.e., that certain structural motifs cannot be directly penalized), we advocate that the motivation for continuing research in their development for solution by quantum computers is strong. We firmly believe that quantum computing is with the potential to increase the tractability of the full RNA folding problem including pseudoknots and suggest this pursuit in earnest.

\section*{Acknowledgment}

We thank the NSERC CREATE in Quantum Computing program and the University of Victoria for providing funding for this project. Additionally, we thank the organizers and sponsors of QHack 2022 for offering a platform through which to showcase this work. We especially thank Amazon Braket for awarding us the compute credits to carry out our experiments.

\bibliographystyle{IEEEtran}
\bibliography{IEEEabrv, qRNA}

\newpage
\begin{center}
\textsc{Appendix}
\end{center}
\smallskip
\renewcommand{\theequation}{A.\arabic{equation}}
\setcounter{equation}{0}

\noindent\rule{\columnwidth}{1pt}
\smallskip

The QUBO formulations for RNA folding presented herein as well as that ref. \cite{Fox2021} requires that each candidate stem of a query RNA be enumerated and mapped to a qubit. Further, the pairwise interactions between these candidate stems must also be enumerated. It is therefore useful to consider the upper limits to the number of stems and their pairwise interactions to estimate bounds for both classical pre-processing time-complexity and quantum spatial resource requirements.

We work with the extreme case of an RNA consisting in alternating bases of a canonical base pair along its sequence. For example: \inlinecode{GCGCGC...} or \inlinecode{AUAUAU...}. We denote the sequence length of the query RNA $N$, the minimum stem size $m$, the function over these two variables which yields the number of candidate stems $S(N, m)$, and the number of pairwise interactions $P(N, m)$. 
\smallskip

\noindent\rule{\columnwidth}{1pt}
\smallskip

\noindent Consider a \inlinecode{GCGCGC...} sequence in the matrix formation of Figure 1, and suppose odd $N$. Then, there are $N-1$ anti-diagonals containing all $1$s and $N-2$ diagonals containing all $0$s. We refer to the $1$-containing anti-diagonals as ``anti-diagonals'' from here.

\smallskip
\smallskip

\noindent By symmetry, $(N-1)/2$ of these anti-diagonals are with unique lengths, these lengths following the sequence: $1, 2, 3, ... , (N-1)/2$. If stems are limited to being of size $m$ or larger, replace $(N-1)/2$ with $(N-1)/2-m+1$. To find the total number of stems, we count over each anti-diagonal the number of sub-anti-diagonals of length $\geq m$. Allowing $i$ to be the anti-diagonal length and $j$ to be the stem length, both with lower bounds of $m$:

\smallskip
\smallskip

\beq
    S(N, m) = 2\sum_{i=m}^{\frac{N-1}{2}}\sum_{j=m}^{i}(i-j+1)
    \label{SNM}
\eeq

\smallskip
\smallskip

\noindent The summand $i-j+1$ tells simply how many stems of a length $j$ fit within a anti-diagonal of length $i$.

\smallskip
\smallskip

\noindent Now, evaluating the inner sum, we find:

\beq
    \sum_{j=m}^{i}(i-j+1) = \frac{1}{2}(i-m+1)(i-m+2)
\eeq

\noindent Therefore, A.1 becomes with this substitution:

\beq
    S(N, m) = \sum_{i=m}^{\frac{N-1}{2}}(i-m+1)(i-m+2)
\eeq

\noindent This then evaluates to:

\beq
    S(N, m) = \sum_{k=1}^{n}k(k+1) = \frac{1}{3}n(n+1)(n+2)
\eeq 

\noindent Where $k = i-m+1$ and $n = \frac{N-1}{2}-m+1$. Resubstituting, we find the analytic expression:

\beq
\begin{split}
    S(N, m) = \frac{1}{24}\Big(&N^3+N^2(9-6m)\\
    &+N(12m^2-36m+23)\\
    &+(-8m^3+36m^2-46m)+15\Big)
\end{split}
\eeq

\noindent For $N\gg m$, this is $\mathcal{O}(N^3)$, which is to say that the number of stems mapping directly to qubits is cubic in sequence length, in the worst case. 

\smallskip
\smallskip

\noindent Now, given that every stem may form a pseudoknot/overlap with any other stem, we enumerate the total number of possible pairwise interactions between stems we must check as:

\beq
    P(N, m) = \binom{S(N, m)}{2} = \frac{S(N,m)(S(N,m)-1)}{2}
    \label{PNM}
\eeq

\noindent In the $N\gg m$ approximation, this is $\mathcal{O}(N^9)$, such that the number of pairwise interaction  grows to the ninth power of sequence length, in the worst case.

\smallskip
\smallskip

\noindent Having discussed the space-complexity of our pre-processing, we now consider the time-complexity, which again divides into two stages: enumerating over potential stems and enumerating pairwise interactions between stems. 

\smallskip
\smallskip

\noindent To find stems, we iterate over all upper anti-diagonals of the RNA potential stems matrix. Straightforwardly, by considering all $1$s as beginnings to potential stems and maintaining these in a stack until proven otherwise, we achieve a full enumeration in $\mathcal{O}(N^2)$ time, again where $N\gg m$.

\smallskip
\smallskip

\noindent Once we are with a list of stems, finding the overlaps and pseudoknots of a set of stems is precisely an interval overlap problem, which may be solved in $\mathcal{O}(S(N,m)\log S(N,m)) = \mathcal{O}(N^3\log N)$ time using interval tree methods. 

\smallskip
\smallskip

\noindent In sum, with the worst-case RNA, the number of potential stems is $\mathcal{O}(N^3)$, requiring $\mathcal{O}(N^2)$ time to find, and the number of pairwise interactions between potential stems is $\mathcal{O}(N^9)$, requiring $\mathcal{O}(N^3\log N)$ time to find. This grants us a final pre-processing time-complexity of $\mathcal{O}(N^3\log N)$ and a quantum space-complexity of $\mathcal{O}(N^3)$ (under the assumption of full connectivity of the qubit graph; in reality, extra qubits are required to achieve proper coupling, at least in the case of quantum annealing).

\smallskip
\smallskip

\noindent For the refs. \cite{Lewis2021, Lewis2021a} QUBO, only stems of length 2 ($m = 2$) are considered. Therefore, equation \ref{SNM} reduces to $\mathcal{O}(N^2)$ in the worst case, and equation \ref{PNM} reduces to $\mathcal{O}(N^4)$. The pairwise interaction terms can therefore be computed in $\mathcal{O}(N^2\log N)$ time, such that the pre-processing stage of this algorithm is of $\mathcal{O}(N^2\log N)$ time-complexity, and a quantum space-complexity of $\mathcal{O}(N^2)$ is required in the worst case.

\end{document}